\documentstyle[12pt,epsfig]{article}
\begin{document}
\begin{center}
{\Large On Neutrino-Mixing-Generated Lepton Asymmetry\\ 
and the Primordial Helium-4 Abundance}\\
\ \\
M. V. Chizhov\footnote{Permanent address: Centre for Space Research 
and Technologies,
Faculty of Physics,\\ University of Sofia, 1164 Sofia, Bulgaria}
and D. P. Kirilova\footnote{Permanent address: Institute of Astronomy, 
Bulgarian Academy of Sciences,\\
blvd. Tsarigradsko Shosse 72, 1784 Sofia, Bulgaria}\\
\ \\
{\it The Abdus Salam International Centre for Theoretical Physics,\\ 
Strada Costiera 11, 34014 Trieste, Italy}
\end{center}

\begin{abstract}
In this article  we discuss lepton asymmetry effect on BBN 
with neutrino oscillations. We argue that
{\it asymmetry much smaller than 0.01},
although not big enough to influence directly the nucleosynthesis 
kinetics, {\it can  effect considerably BBN indirectly via neutrino
oscillations}. Namely, it distorts 
neutrino spectrum and changes  neutrino density evolution 
 and  the pattern of oscillations 
(either suppressing or enhancing them), which in turn effect 
the primordial synthesis of elements. We show that the results 
of the paper X. Shi et al., Phys. Rev. D 60, 063002 (1999), 
 based on the assumption that only $L>0.01$ will
influence helium-4 production, are not valid. 
Instead, the precise
constraints on neutrino mixing parameters from BBN are presented. 
\end{abstract}

   There exists an interesting interplay between lepton asymmetry
and neutrino oscillations in the early Universe. As it was noticed in
\cite{BD90,FTV96,Shi96,NU96} neutrino
oscillations can generate lepton asymmetry, besides their well known
ability to erase it \cite{BD,Enqvist,FV95}. On the other hand
lepton asymmetry (no matter if neutrino-mixing generated or pre-existing one)
can suppress neutrino oscillations~\cite{FV95,NP} 
and  has also the remarkable ability to enhance them~\cite{NP}.
Consequently, in the presence of neutrino oscillations, lepton
asymmetry exerts much complex influence on Big Bang Nucleosynthesis 
 (BBN) via oscillations,   
than  in the simple case without oscillations.

In this work we will  discuss the indirect effect of lepton
asymmetry on primordial nucleosynthesis via  neutrino 
oscillations. This paper is provoked by the publication
``Neutrino-Mixing-Generated Lepton Asymmetry and the Primordial He-4
Abundance" by X. Shi, G. Fuller, and K. Abazajian, published in 
Phys. Rev. D 60, 063002 (1999) ref.~\cite{SFA} (hereafter SFA). 
As we understood from their  paper and some other recent
publications~\cite{sorry, wrong, Shi96} there exists some  
shallow understanding of the role of the lepton asymmetry in 
BBN with oscillations. And we would like on the first place to 
clarify this subject. 

In SFA the study of the lepton asymmetry effect on BBN is
based on the assumption that only asymmetry bigger than 0.01 at the
freeze-out of the $n-p$ transitions may have an appreciable impact 
on the primordial
abundance of helium-4 $Y_p$. Hence, the authors estimate the effect of 
the asymmetry on BBN {\it after} it has been enhanced up to 
0.01. 
 Certainly such consideration is valid for the simple case of nucleosynthesis
{\it without oscillations}! There are exhaustive
studies on that subject~\cite{lnuc}, which results,
concerning neutrino degeneracy effect on nucleosynthesis, the authors 
of SFA reproduce in general. 

However, in the case of nucleosynthesis {\it with oscillations} 
the assumption that only asymmetry bigger than 0.01 effects
nucleosynthesis, is no longer valid. It was first noticed in  
the original works~\cite{NU96,NP},
that in the case of BBN with neutrino oscillations 
even a very small lepton asymmetries $L<<0.01$ (either initially
present~\cite{NP}, or 
dynamically `neutrino-mixing' generated~\cite{NU96}),
 although not big
enough to influence nucleosynthesis directly,  may 
considerably effect 
BBN  indirectly through oscillations. 
In these works a very precise account of the evolution of the  
 neutrino and antineutrino distribution functions and 
their spectral distortions, and the evolution of the asymmetry 
was provided in the BBN calculations.\footnote{We are really sorry that 
the authors of SFA had to rediscover the importance of this account, 
but we cannot agree neither that they were the first to provide the account, 
nor that they provided this account accurately.}

In the present work we calculate the net effect of small
lepton asymmetries $L << 0.01$ on BBN and obtain precise 
cosmological constraints on neutrino mixing parameters. 

In the presence of oscillations,
lepton asymmetry affects BBN {\it indirectly} through
its feedback effect on:\\
(1) the evolution of the neutrino and antineutrino number
 densities~\cite{Enqvist,NU96}, which play an essential role in the 
kinetics of nucleons at $n/p$-freeze-out;\\
(2) the neutrino and antineutrino spectrum distortion~\cite{NU96,NP}, 
which is important for the correct calculation of the neutrino 
number densities and weak interaction rates in $n-p$ transitions
(see the following eq.~(2));\\
(3) the neutrino oscillation pattern. Namely, $L$ may suppress or enhance
oscillations, leading, correspondingly, to underproduction or 
overproduction of primordial helium-4 in comparison with the case 
without asymmetry account. 
 The suppression may be  strong enough to 
allow substantial alleviation of the nucleosynthesis bounds on the neutrino 
mixing parameters. The effect on BBN of a suppression due to a
{\it relic} neutrino asymmetry 
was discussed first in \cite{FV95} and calculated in detail with 
the account of (1) and (2) in \cite{NP}. While the suppression 
due to {\it neutrino-mixing generated} asymmetry and its effect on BBN 
was calculated first in \cite{NU96}. It was
recently shown~\cite{NP} also that  
lepton asymmetry is capable of enhancing the oscillations and thus 
strengthening of the BBN bounds on the neutrino oscillation parameters.
 
These three effects are typical for the case of BBN {\it with
oscillations}. 

(4) In case when $L$ is~\cite{lnuc} or grows~\cite{FVlast} big enough 
$> 0.01$ it can also influence 
{\it directly} the kinetics of the $n-p$ transitions, depending 
on the sign of $L$. 

It is essential that 
in the presence of oscillations lepton asymmetry has a more complex
influence on BBN (1)-(4) than in the simple case without oscillations.  
The correct study should follow {\it selfconsistently} the evolution 
of the neutrino ensembles, the evolution of the lepton asymmetry 
as well as the evolution of the
neutron and proton number densities.
So that the complete effect of the asymmetry throughout its evolution 
(growth or damping) during the nucleosynthesis epoch could be registered. 
Such an exact study was provided for small neutrino mass differences 
$\delta m^2 \le 10^{-7}$ eV$^2$  for the resonant case in \cite{NU96} 
and in the nonresonant case in \cite{NP,PR}.

In what follows we present the results of a precise investigation 
of the asymmetry effect on BBN via neutrino oscillations and  provide a 
comparison with an artificial case without the account of asymmetry 
in order to extract the net effect
of the asymmetry on BBN. Finally, we obtain accurate cosmological 
constraints on the oscillation parameters.  

We discuss the case of active-sterile neutrino oscillations
assuming mixing present just in 
the electron sector $\nu_i=U_{il}~\nu_l$ ($l=e,s$), 
following the line of work in ref.~\cite{NU96}. 
The set of kinetic equations describing simultaneously 
the evolution of the  neutrino and antineutrino density matrix 
$\rho$ and  $\bar{\rho}$  and the
evolution of the neutron
number density   $n_n$ in momentum space reads:

\begin{eqnarray}
&&{\partial \rho(t) \over \partial t} =
H p_\nu~ {\partial \rho(t) \over \partial p_\nu} +
\nonumber\\
&&+ i \left[ {\cal H}_o, \rho(t) \right]
+i \sqrt{2} G_F \left(\pm {\cal L} - Q/M_W^2 \right)N_\gamma
\left[ \alpha, \rho(t) \right]
+ {\rm O}\left(G_F^2 \right),
\label{kin}
\end{eqnarray}
\begin{eqnarray}
&&\left(\partial n_n / \partial t \right)
 = H p_n~ \left(\partial n_n / \partial p_n \right) +
\nonumber\\
&& + \int {\rm d}\Omega(e^-,p,\nu) |{\cal A}(e^- p\to\nu n)|^2
\left[n_{e^-} n_p (1-\rho_{LL}) - n_n \rho_{LL} (1-n_{e^-})\right]
\nonumber\\
&& - \int {\rm d}\Omega(e^+,p,\tilde{\nu}) |{\cal A}(e^+n\to
p\tilde{\nu})|^2
\left[n_{e^+} n_n (1-\bar{\rho}_{LL}) - n_p \bar{\rho}_{LL} (1-n_{e^+})\right].
\end{eqnarray}
where $\alpha_{ij}=U^*_{ie} U_{je}$,
$p_\nu$ is the momentum of electron neutrino,
 $n$ stands for the number density of the interacting particles,
${\rm d}\Omega(i,j,k)$ is a phase space factor and  ${\cal A}$ is the
amplitude of the corresponding process. 
The sign plus in front of ${\cal L}$ corresponds to neutrino
ensemble, while minus - to antineutrino ensemble.
Actually, we solve nine equations selfconsistently:
four equations for the components of
the neutrino density matrix,  another four for
 the  antineutrino density matrix following
from eq.~(1), and one for the neutron number density eq.~(2).

The first term in the right hand side of the equations (1) 
and (2)  describes the effect of
Universe expansion.
The second term in (1) is responsible for neutrino oscillations, the
third accounts for forward neutrino scattering off the
medium and the last one accounts for second order interaction
effects of neutrinos with the medium.   
${\cal H}_o$ is the free neutrino Hamiltonian.
${\cal L}$ is proportional to the fermion asymmetry of the plasma 
and is essentially expressed through the neutrino asymmetries
${\cal L} \sim 2L_{\nu_e}+L_{\nu_\mu}+L_{\nu_\tau}$,
where
$L_{\mu,\tau} \sim (N_{\mu,\tau}-N_{\bar{\mu},\bar{\tau}})/ N_\gamma$
and $L_{\nu_e} \sim \int {\rm d}^3p (\rho_{LL}-\bar{\rho}_{LL})/N_\gamma$.
The `nonlocal' term $Q$ arises as an $W/Z$ propagator effect,
$Q \sim E_\nu~T$.
It is important for the nonequilibrium active-sterile neutrino oscillations
to provide simultaneous account of the different competing processes,
namely: neutrino oscillations, Hubble expansion and weak interaction processes.

Neutrino and antineutrino ensembles evolve differently as far as the
background is not $CP$ symmetric.
Besides, the evolution of neutrino and antineutrino ensembles may become 
strongly coupled due to the growing electron asymmetry term and hence, 
the evolution of  $\rho$ and $\bar{\rho}$  
must be considered simultaneously.

Moreover, it is extremely important for the correct account of the role of the
asymmetry on BBN to study  
the asymmetry evolution and the neutron number density
evolution in $p$-space selfconsistently with 
the evolution of neutrino and
antineutrino ensembles involved in
oscillations! This looks obvious as far as  there
exists asymmetry-oscillations interplay -- 
oscillations  change
neutrino-antineutrino asymmetry and it in turn affects oscillations,
and, besides, neutrino  $\rho_{LL}$ and antineutrino $\bar{\rho}_{LL}$ number
densities enter the
kinetic equations for nucleons. However,
usually in many papers the growth of asymmetry is calculated, and 
then, when it  has reached values around $0.01$, its influence on BBN 
kinetics is estimated. Thus, the asymmetry influence (1)-(3) on 
BBN during its growth till $0.01$ cannot be caught.  
We will demonstrate in this work that this very influence may give up 
to $10\%$ relative change in primordial helium-4.    
Therefore, the indirect influence of lepton asymmetry on BBN should be 
carefully accounted for during asymmetry's full  evolution.

It is essential also, that the equations should follow neutrino evolution 
in momentum space, i.e. enabling to account precisely
for the distortion of the neutrino
spectrum due to  oscillations and asymmetry. This approach was
demonstrated~\cite{NU96,PR} in detail for the case of
small mass differences and it helped to precise the constraints on the
neutrino squared mass differences $\delta m^2$ by almost an order of magnitude in
comparison with the previous studies (see fig.~8 from~\cite{PR}).
Working with mean energies and 
equilibrium spectrum is tempting of course due to the simplicity of the
analysis, however, is not correct. 
 We have stressed  in our previous works the importance of the proper
account of the spectrum distortion
and asymmetry for the BBN with active-sterile oscillations and we have
provided this account in \cite{NU96, PR, NP}. Besides, many papers  
have discussed separately the  questions of the dynamical evolution
of the asymmetry (see for example ~\cite{Enqvist,FTV96,FV95,FVlast})  
or of the correct account of 
the spectrum distortion for nonequilibrium 
neutrinos~\cite{D81,D98}.\footnote{Therefore, it
is quite amazing now in 1999 to see published statement in (SFA) 
about the existence in
literature only of ``BBN calculations based on a constant asymmetry and a
thermal neutrino spectrum" ``overly simplistic" and with ``inaccurate
results". It is easy to judge  that this same paper (SFA)
 may be considered overly simplistic
comparing the calculated in it ``semianalytically" distorted
spectrum distributions with the precisely calculated 
 spectra presented in \cite{NU96}.}

It is really not an easy task to solve exactly the system of 
eqs.~(1)-(2).
Especially,  in the case of a  rapid asymmetry
growth   more than 1000 bins may be required  
for the accurate description of the neutrino spectrum, 
however, this  is the correct way to study this topic.
We have described the spectrum using in general 1000 bins, and  sometimes  
for the resonant case up to 5000 bins. The equations were integrated for the 
characteristic period 
from the electron neutrino decoupling  till the $n/p$ freeze-out at 0.3 MeV. 
We have calculated the value of the 
primordially produced helium-4 with neutrino  oscillations for  
the full range of the model's parameters values, namely 
for $\sin^2(2\theta)$ ranging from  $10^{-3}$ to maximal mixing 
and $\delta m^2 \le 10^{-7}$ $eV^2$. 
For smaller mixing parameters the effect on
helium-4 is negligible~\cite{NU96}. 
The exact feedback effect of the asymmetry on the neutrino ensembles evolution, 
neutrino spectrum distortion and neutrino oscillations, was  numerically 
followed. Hence, the total effect of the asymmetry on BBN, indirect via its
interplay with oscillations and direct on the kinetics of $n-p$ transitions 
was obtained numerically. 

In fig.~1 the impact in helium-4  due to oscillations and asymmetry 
is presented as a function of the
 neutrino square mass differences. For comparison the curve corresponding to the
artificial case without the account of the asymmetry is presented also. 
The difference between the two curves measures the net asymmetry effect 
on BBN with oscillations. It is obvious, that for the range of 
oscillation parameters discussed, the total effect of the asymmetry 
is a reduction in $Y_p$ in comparison with the case without asymmetry account. 
This reduction  can be as big as $10\%$,  which is 
considerable on the background of our recent knowledge from  
primordial helium measurements~\cite{BBN}.
As it is obvious from the figure, small $\delta m^2$ are also constrained 
from BBN considerations. The obtained constraints on $\delta m^2$
are by {\it several orders of magnitude} more severe
than the constraints obtained in SFA (see fig.4 there\footnote{or 
the same figure reproduced in other publication of the same authors, 
namely fig.3 in the first reference in \cite{wrong}}) on the
basis  only of the kinetic effect of the asymmetry.

In fig.~2 we present a comparison of the iso-helium-4 contours,
$Y_p=0.245$, 
for the resonant case, obtained without the account of the asymmetry,   
with the contours obtained with the account of the 
asymmetry.  The area to the left of the curves is 
the allowed region of the oscillation parameters.
 
The numerical analysis showed that in  the case of 
small mass differences we
discuss and naturally small initial asymmetry, 
the growth of the asymmetry is less than 4 orders of magnitude. Hence,
beginning with asymmetries of the order of the baryon one, 
the asymmetry does not grow enough
to influence directly $n-p$ transitions.   
 Consequently, the apparently great
asymmetry effect (as seen from the
curves) is totally due to the indirect effects (1-3) of the asymmetry
on BBN.
The maximal asymmetry effect is around $10\%$ 'underproduction' of 
$Y_p$ in comparison with the case of BBN with oscillations but 
without the asymmetry account. 
The total effect of oscillations, with the complete 
account of the asymmetry effects, is still overproduction of 
helium-4, although considerably smaller 
than in the calculations neglecting asymmetry. Therefore,  
nucleosynthesis constraints on the mixing parameters of neutrino 
are alleviated considerably due to the asymmetry effect.      

The case of {\it nonresonant} active-sterile oscillations was already 
discussed~\cite{NU96} and investigated in detail in \cite{PR}.
It was shown that the effect of the 
asymmetry on BBN with oscillations, in case it was initially of the order of the
baryon one, is negligible. However, in case it was initially bigger than
$10^{-7}$, it may have also crucial effect on BBN through its effect on
oscillations~\cite{NP}. In the last work a complete exact 
numerical study of the asymmetry effect on BBN with oscillations was provided for a
wide range of
initial asymmetry values ($10^{-10} - 10^{-2}$). 
In fig.~4 of the original paper~\cite{NP} 
 the iso-helium contours for the case with 
pre-existing asymmetry ($L=10^{-6}$) and the case without asymmetry effect (dashed
curves) were presented. It is obvious that in the discussed nonresonant case 
the strong asymmetry effect again is due to its indirect influence on
nucleosynthesis. 
However, it is not so straightforward. For the nonresonant case the asymmetry
account reflects into
alleviating BBN constraints on mixing parameters for big $\theta$ 
due to suppression of oscillations but 
strengthening the constraints for small $\theta$ due to 
enhancement of oscillations.
For more details see the original paper \cite{NP}. 

 We would like also to stress, that in the nonresonant 
case of small mass differences
oscillations, due to the complex interplay 
between oscillations and asymmetry,  antineutrinos and
neutrinos undergo  resonance   almost simultaneously. 
This is easy to grasp  as far as the asymmetry has a fast oscillating
sign-changing behavior due to which both the neutrino and antineutrino
ensembles are able to experience resonance. Consequently, 
the  effect on helium-4 does not depend on the initial sign of 
$L$,  in case the asymmetry is small enough not to have a direct 
kinetic effect on $n-p$ transitions, i.e. $L<< 10^{-2}$ 
(contrary to the case of direct $L$ influence when the sign of $L$ 
is important, as far as in one case it leads to overproduction and in the other to
underproduction of helium-4~\cite{lnuc}.)

 And at last but not least, we are really amused, that the authors 
of SFA after
frankly declaring that they do not know if the  evolution of the lepton
asymmetry represents a true chaos or not,  do continue working with 
this not clear understanding,  
and, moreover, they continue exploiting it fabricating models 
and constraints~\cite{domain}, 
before clarifying the situation with the ``chaotic" behavior of $L$. 

In this work we have proved that lepton asymmetry,
by orders of magnitude less than $0.01$, although not big
enough to influence nucleosynthesis directly, can considerably effect 
nucleosynthesis indirectly via oscillations,
changing the pattern of neutrino  oscillations,
neutrino densities evolution and neutrino spectrum. In the resonant case 
we have obtained precise cosmological constraint on neutrino oscillation 
parameters $\delta m^2$ and $\theta$ accounting  for the dynamical 
evolution of the neutrino asymmetry, its interplay with oscillations 
and its effect on primordial production of helium-4.~\footnote{ The 
constraints for the nonresonant case were obtained in our previous 
work~\cite{PR}.}

The constraints and conclusions of previous works 
\cite{Shi96,SFA,sorry,wrong,FV95,FVlast,domain} concerning 
asymmetry effect on BBN with oscillations will change  considerably  
when a proper selfconsistent account for 
(a) the complete effect of the 
asymmetry (1)-(4) during its whole evolution in nucleosynthesis epoch;
(b) the neutrino spectrum distortion; 
(c) and the exact kinetics of  nucleons  
is provided using the kinetic equations in momentum space. 
 The role of the
mixing-generated neutrino asymmetry 
in BBN is considerable and should be accounted for precisely.
 
We are thankful to ICTP, Trieste, for the financial help and 
hospitality during the preparation of this work. D.K. is 
grateful to prof. Sciama for the participation 
into the astrophysics program this summer and for 
useful discussions.

\pagebreak[1]

\newpage

{\bf Figure captions}\\

{\bf Figure 1.} The relative change in the primordial yield of helium-4 
as a function of the neutrino squared mass differences in case of 
BBN with oscillations for $\sin^2(2\theta)=0.05$. 
The solid curve shows the complete effect 
of oscillations with the account of the asymmetry. The dashed curve 
shows solely the effect of oscillations neglecting the asymmetry.\\

{\bf Figure 2.} On the $\delta m^2-\theta$ plane iso-helium-4 
contour $Y_p=0.245$,
calculated in the discussed model of BBN with active-sterile 
neutrino oscillations
and the account of the complete asymmetry effect, is shown.   
The dashed curve presents a comparison with the same case, but 
without the asymmetry account. The area to the left of the curves is
the allowed region of the oscillation parameters. 
 
\begin{figure}
\epsfig{file=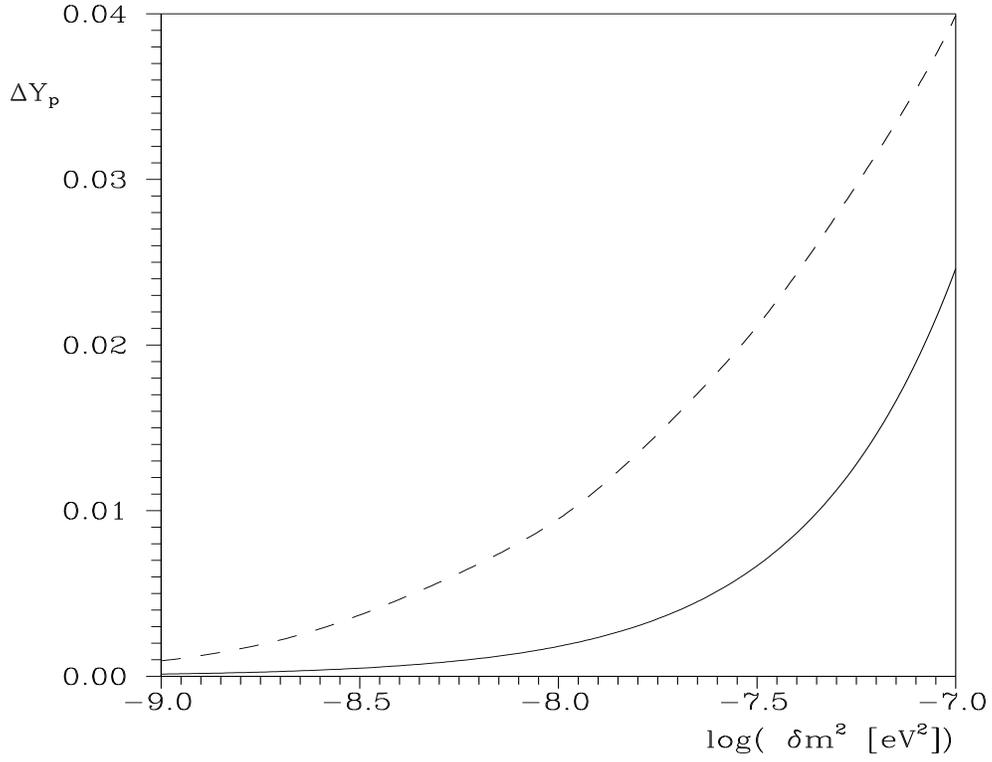,height=10cm,width=13cm}
\caption{The relative change in the primordial yield of helium-4   
as a function of the neutrino squared mass differences in case of
BBN with oscillations for $\sin^2(2\theta)=0.05$.
The solid curve shows the complete effect
of oscillations with the account of the asymmetry. The dashed curve
shows solely the effect of oscillations neglecting the asymmetry.}
\end{figure}

\begin{figure}
\epsfig{file=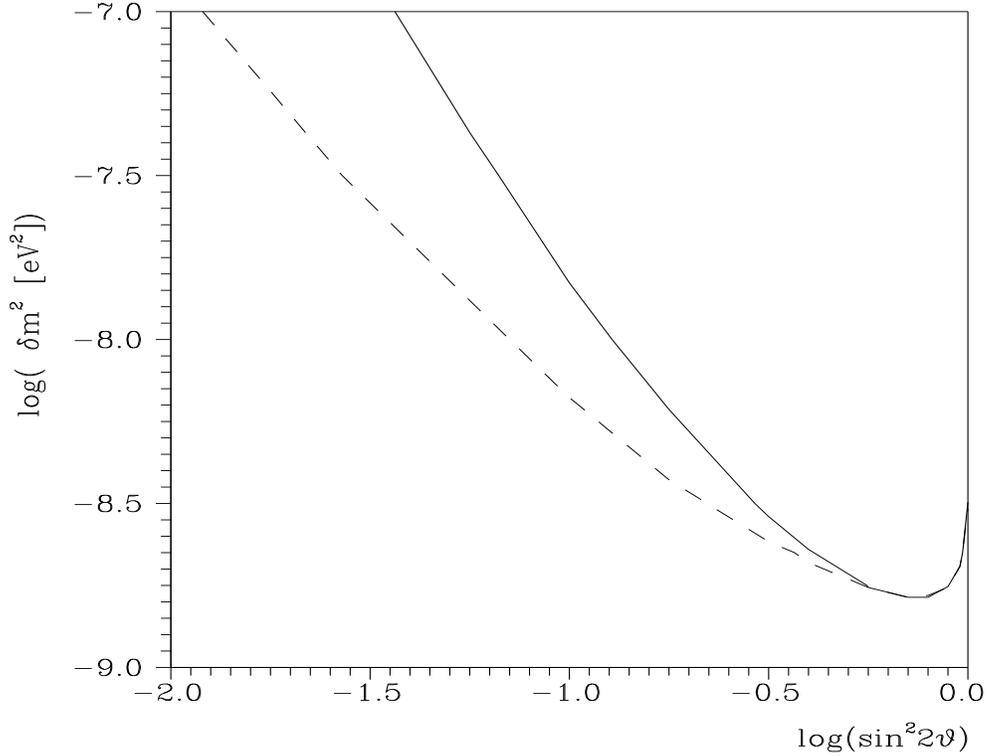,height=10cm,width=13cm}
\caption{On the $\delta m^2-\theta$ plane iso-helium-4
contour $Y_p=0.245$,
calculated in the discussed model of BBN with active-sterile
neutrino oscillations
and the account of the complete asymmetry effect, is shown.
The dashed curve presents a comparison with the same case, but
without the asymmetry account. The area to the left of the curves is
the allowed region of the oscillation parameters.}
\end{figure}

\end{document}